\documentclass[conference]{IEEEtran}
\IEEEoverridecommandlockouts
\usepackage{cite}
\usepackage{amsmath,amssymb,amsfonts}
\usepackage{algorithmic}
\usepackage{graphicx}
\usepackage{textcomp}
\usepackage{listings}
\usepackage{svg}
\usepackage{xcolor}

\usepackage[colorinlistoftodos,textsize=scriptsize]{todonotes}

\lstdefinestyle{smallpython}{
    language=Python,
    basicstyle=\footnotesize\ttfamily,
    showstringspaces=false,
    framerule=0.4pt,
    breaklines=true,
    framesep=3pt,
    aboveskip=6pt,
    belowskip=4pt
}

\def\BibTeX{{\rm B\kern-.05em{\sc i\kern-.025em b}\kern-.08em
    T\kern-.1667em\lower.7ex\hbox{E}\kern-.125emX}}
\begin{document}

\title{Analyzing the Impact of Participant Failures in Cross-Silo Federated Learning \\
\thanks{This work was co-funded by the Federal Ministry of Research, Technology and Space (BMFTR) under Grant 13FH587KX1 (FederatedForecasts)
\\
© 2025 IEEE. Personal use of this material is permitted. Permission from IEEE must be
obtained for all other uses, in any current or future media, including
reprinting/republishing this material for advertising or promotional purposes, creating new
collective works, for resale or redistribution to servers or lists, or reuse of any copyrighted
component of this work in other works.}
}

\author{\IEEEauthorblockN{Fabian Stricker}
\IEEEauthorblockA{%
\textit{Institute of Data-Centric}\\ 
\textit{Software Systems (IDSS)},\\
\textit{Hochschule Karlsruhe --}\\ 
\textit{University of Applied Sciences},\\
Karlsruhe, Germany, and\\
\textit{Technische Universität Berlin},\\
Berlin, Germany\\
fabian.stricker@h-ka.de}
\and
\IEEEauthorblockN{David Bermbach}
\IEEEauthorblockA{\textit{Scalable Software Systems} \\ \textit{Research Group} \\
\textit{Technische Universität Berlin}\\
Berlin, Germany \\
db@3s.tu-berlin.de}
\and
\IEEEauthorblockN{Christian Zirpins}
\IEEEauthorblockA{%
\textit{Institute of Data-Centric}\\ 
\textit{Software Systems (IDSS)},\\
\textit{Hochschule Karlsruhe --}\\ 
\textit{University of Applied Sciences},\\
Karlsruhe, Germany\\
christian.zirpins@h-ka.de}
}

\maketitle

\begin{abstract}
Federated learning (FL) is a new paradigm for training machine learning (ML) models without sharing data.
While applying FL in cross-silo scenarios, where organizations collaborate, it is necessary that the FL system is reliable; however, participants can fail due to various reasons (e.g., communication issues or misconfigurations).
In order to provide a reliable system, it is necessary to analyze the impact of participant failures.
While this problem received attention in cross-device FL where mobile devices with limited resources participate, there is comparatively little research in cross-silo FL.

Therefore, we conduct an extensive study for analyzing the impact of participant failures on the model quality in the context of inter-organizational cross-silo FL with few participants.
In our study, we focus on analyzing generally influential factors such as the impact of the timing and the data as well as the impact on the evaluation, which is important for deciding, if the model should be deployed.
We show that under high skews the evaluation is optimistic and hides the real impact.
Furthermore, we demonstrate that the timing impacts the quality of the trained model.
Our results offer insights for researchers and software architects aiming to build robust FL systems.
\end{abstract}


\section{Introduction}
% Introduction
%%Short intro - FL, Process (local data, use global model, goal of the participants)
Federated learning (FL) is a paradigm for training a machine learning (ML) model collaboratively without sharing data; hence, improving communication-efficiency and privacy~\cite{mcmahanCommunicationEfficientLearningDeep2017a,paper_stricker_fl-apu}.
In FL, each participant trains an ML model locally with their data and sends the trained model to a coordinator that aggregates the local models into a global model. 
This process is repeated for multiple rounds and as result of the collaboration, each participant receives the global model~\cite{mcmahanCommunicationEfficientLearningDeep2017a,paper_stricker_fl-apu}. 

In FL, each participant has a part of the data; hence, if a participant crashes, it drops out of the process and the available training data change, which influences the quality of the model.
In cross-silo FL, where participants are organizations, failures such as crashes are less likely than in cross-device FL, where mobile devices with limited resources are involved. However, participants can still drop out of the process due to issues such as communication problems, misconfiguration, power outages, or software bugs.
If such a failure occurs, unique and important information can be lost, however, it is not well studied how this affects performance as there is a variety of influence factors.
Consequently, to achieve a reliable FL system and understand the risks of participant failures, it is necessary to analyze and evaluate the impact on the quality.
In cross-silo FL, low model quality due to participant failure would lead to possibly severe losses for all participating organizations.
Consequently, it is important that the FL system is robust and produces a reliable model.
While the dropout of participants has received attention in cross-device FL~\cite{riberoFederatedLearningIntermittent2023a, sousaEnhancingRobustnessFederated2025, imteajSurveyFederatedLearning2022}, such research on cross-silo FL is still very limited.

To bridge this research gap, we conduct an experimentation study, where we first analyze which factors can influence the impact of participants failures and second, provide extensive experiments on the impact of participant failures on the global model performance.
In our study, we consider inter-organizational cross-silo FL scenarios where a small number of companies collaborate by investing resources to jointly train a model, with data being non-independent and -identically distributed (non-IID)~\cite{zhaoFederatedLearningNonIID2018}.
Furthermore, to the best of our knowledge, there is no research on the impact of participant failures during the model evaluation, which is essential for assessing the quality of the model.
We show that participant contribution metrics can indicate the impact if a specific valuable participant drops out and that the timing of the dropout is important. 
If a participant joins in the last two rounds of the FL process, this can improve the quality more than a participant that only contributes in early training rounds as the model forgets and learns information.
In summary, we make the following contributions:
\begin{itemize}
    \item We identify and discuss important influencing factors that determine the impact of failed participants (Section~\ref{sec:influencing_factors}).
    \item We conduct numerous experiments across different factors such as datasets, skews (i.e., quantity and label skews), evaluation, and timing of failure. Furthermore, we analyze if participant contribution metrics can indicate the impact of a dropout and we analyze if the failed participant can use the model for its use case (Section~\ref{sec:exp_plan}).
    \item We provide a thorough discussion of our findings as a guidance for assessing the problem of participant dropouts in cross-silo FL (Section~\ref{sec:findings}).
\end{itemize}
In addition to the structure in the contributions, we discuss required background knowledge in Section~\ref{sec:background_related_work}.

\section{Background \& Related Work}
\label{sec:background_related_work}
This section discusses the background and the key related work on FL participant failures. 

\subsection{Cross-Silo and Cross-Device FL}
In FL, two deployment scenarios are widely adopted: cross-silo and cross-device FL.
In cross-device FL, the participants are often edge- or mobile devices that come in large numbers and have limited resources (i.e., energy, data, hardware); hence, they may not participate in every round and only a fraction is available at a time. 
Due to the large number, communication is a bottleneck and managing all devices is difficult, even more because they are often unreliable. 
In comparison, cross-silo FL usually includes $2-100$ organizational participants. 
Often, these are reliable nodes with plenty of resources that can be efficiently managed \cite[p.4-10]{kairouzAdvancesOpenProblems2021}.

\subsection{Partial Participation in Cross-Device FL}
Cross-device FL considers mobile devices that are unreliable. Here, the likelihood that a participating device will not join in all training rounds is high -- it will rather partially participate in the process, joining and leaving over time.
Partial participation can also be caused by the fact that more participants are available than a single coordinator can efficiently manage.
Consequently, the coordinator has to apply participant selection mechanisms in order to choose a reasonably large subset of participants with diverse and sufficient data~\cite{bonawitzFederatedLearningScale2019,imteajSurveyFederatedLearning2022}. 

\subsection{Participant Contribution}
Measuring the importance of a participant is necessary in order to understand its value for the FL training.
In the current state-of-the-art, there is a variety of methods for quantifying contributions, e.g.,~\cite{ghorbaniDataShapleyEquitable2019,liuGTGShapleyEfficient2022,wangPrincipledApproachData2020}. Most of them build on Shapley values (SV), which were initially used in game theory but are also applicable in FL. Here, the cooperative game can be reformulated to multiple data sources that cooperate and SVs are used to assess how valuable their data sets are~\cite{wangPrincipledApproachData2020}. 
%% A Principled Approach to Data Valuation for Federated Learning (USED SV-Estimation)
Calculating SVs, however, is costly as it requires testing all possible coalitions, which hinders their application for a wide range of real-world settings. 
Instead of retraining with different coalitions of data sources, the model updates can be used to evaluate the importance of a participant~\cite{liuGTGShapleyEfficient2022}. 
Furthermore, Wang et al.~\cite{wangPrincipledApproachData2020} proposed methods that approximate the SVs to reduce the overhead. 

\subsection{Related Work}
While there is research on missing participants in cross-device FL, this is much less the case for cross-silo FL. 

There are a few works that focus on participant selection methods to mitigate the impact of unavailable participants.
Ribero et al.~\cite{riberoFederatedLearningIntermittent2023a} proposed an unbiased selection strategy that includes the availability of participants to prevent bias.
Sousa et al.~\cite{sousaEnhancingRobustnessFederated2025} propose an add-on for selection strategies that uses a reserve pool of substitute participants to ensure convergence of the global model in case of participant failures.
However, using a reserve pool only is applicable if there are sufficient participants.

Besides participant selection strategies, there are also substitution and augmentation methods to tackle convergence issues~\cite{xuStabilizingImprovingFederated2025}.
Wang and Xu~\cite{wangFriendsHelpSaving2024} use the model update of a participant that has a similar data distribution as substitute for the update of the failed participant to reduce the impact. Their approach is based on similarity and clustering which is difficult to achieve with few participants and skewed data. Instead, of using a similar participant as substitution, Sun et al.~\cite{sunMimiCCombatingClient2024} modify model updates with a correction value to mitigate convergence issues in mobile edge networks.

Besides mitigating the issues, there are also works that focus on attacking FL with dropout attacks.
Qian et al.~\cite{qianDROPFLClientDropout2024a} propose a dropout attack under constraints such as low bandwidth and unstable network connection. They found that attacking a high value participant slows down the training process and has more influence in a non-IID setting.

One work that partially matches our scenario with only few participants, is conducted by Huang et al.~\cite{huangKeepItSimple2023}. They analyze the impact of unreliable devices in a rural environment and consider non-IID data in combination with different participation rates. Their results show that FedAvg~\cite{mcmahanCommunicationEfficientLearningDeep2017a} is robust against unreliable participants; however, their data are not highly skewed.

% Why is my contribution novel (also only 2-4 sentences)
While some of the related work analyzes the impact of participant failures, all focus on cross-device scenarios.
In comparison, we consider $4-10$ participants and a high label skew. Hence, approaches such as participant selection, clustering, or a reserve pool of participants are not applicable.
However, none of the mentioned works solely focus on analyzing the impact of participant failures in cross-silo FL. 
Consequently, a thorough experimentation study is needed that considers factors such as evaluation or timing.

\section{Influencing Factors}
\label{sec:influencing_factors}
Previous research with focus on participant dropouts has analyzed different aspects such as participation rate with random dropout, aggregation methods, availability patterns and the skewness of data~\cite{qianDROPFLClientDropout2024a,sunMimiCCombatingClient2024,huangKeepItSimple2023}.
Besides these aspects, we want to provide a broad overview of potential factors that can influence the impact of a participant failure in inter-organizational cross-silo FL and discuss why they are influential. This can help in understanding and assessing the potential impact if a specific participant fails. We categorize these aspects into dataset-specific, training-specific, and coordination-specific.

\paragraph{Dataset-specific Factors}
As dataset-specific factors, we consider data skewness and data complexity.
For data skewness, a higher skew (both quantity and label skew) leads to participants with unique data; hence, specific data are only represented by single silos which in case of a dropout would lead to the missing of these unique data (i.e., unique classes or edge cases)~\cite[p.18-20]{kairouzAdvancesOpenProblems2021}. 
The dataset complexity describes the difficulty for a model to extract meaningful patterns from the available data.
This difficulty can increase due to factors such as high feature dimensionality, number of unique labels, and quality of the data. 
In addition, the complexity influences the skewness (i.e., a low number of classes makes it difficult for each participant to have unique labels) as well as the minimal required data for the model to understand a pattern.

\paragraph{Training-specific Factors}
As training-specific factors, we consider the model architecture, hyperparameters (i.e., learning rate, optimizers), and aggregation methods.
If the selected model architecture is suitable for the task, we assume that training an architecture that focuses on generalizing well on out-of-distribution data can reduce the immediate impact of a participant dropout as it is able to perform well on data that it did not train on~\cite{liuOutOfDistributionGeneralizationSurvey2023a}.
Regarding the immediate impact, we assume that if a relevant participant fails it does not immediately make the model unable to predict missing labels as the context and rather the information will be forgotten over the next training rounds. One reason for this assumption is that each training step is built upon the previous one and the global model from the previous round still holds the information.
In this regard, the influencing factors that cause the model weights to change are the learning rate, the optimizer which can also affect the learning rate (i.e., Adam~\cite{kingmaAdamMethodStochastic2015}), and the aggregation method. 
Consequently, these three factors can have potential influence on the impact of participant failures.

\paragraph{Coordination-specific Factors}
As coordination-specific factors, we consider the current task (i.e., training or evaluation) and the participation rate that we split into the duration and the timing of the dropout.
These factors describe the behavior of a dropout and are therefore closely related to its impact on the final global model.

\section{Experiment Plan}
% Experiment Plan
\label{sec:exp_plan}
In this section, we present the goals and the design of our experimentation study, and describe how the experiments are conducted.

\subsection{Experimentation Goals}
We want to explore how the influencing factors in Section~\ref{sec:influencing_factors} affect the quality of the trained model in cross-silo and inter-organizational FL scenario with few participants.
Unfortunately, due to the number of factors and the resulting experiment combinations, it is not feasible to study all of them. 
Therefore, we will focus on a selected set of factors, that we assume to have the highest impact on the model quality.

First, we will analyze if the model evaluation is unaffected by the dropout, otherwise this can lead to wrong assumptions about the quality of the global model.
Second, we examine data skewness, data complexity and number of participants as these combined affect the data distribution, which is the foundation for the quality of the model.
Last, we analyze the timing of the failure and, hence, the availability of a participant.
This is important as it impacts when data are available or unavailable for the FL process.

Besides, these factors, we analyze aspects that are important for cross-silo FL.
In this regard, all participants should benefit from participating. 
Hence, it is necessary to analyze if the failed participant can use the trained model for its application without the need to put in additional effort.
Furthermore, it is important to examine if contribution measures can help in predicting the impact of a selected participant failing during the FL process.

%% Experiment Design
\subsection{Experiment Design}
% Approach
In general, testing all possible combinations for influencing factors would lead to an extensive number of possible experiments. 
For the sake of feasibility, we focus on two base-configurations (see Table~\ref{tab:base_configs}) that align with our scenario.

% Client Contribution Method
\subsubsection{Contribution Measurement}
For quantifying the contribution of a participant, we use \textit{RoundSVEstimation} based on permutation sampling proposed by Wang et al.~\cite{wangPrincipledApproachData2020}. 
We use this method because Shapley-based methods are accurate; however, they also cause high overhead.
To mitigate this problem, the permutation-based method does not require retraining and does not test all possible coalitions.

\subsubsection{Datasets}
For experimentation, we use CIFAR-10~\cite{krizhevskyLearningMultipleLayers2009a} and CIFAR-100~\cite{krizhevskyLearningMultipleLayers2009a} as datasets.
They have a similar structure but feature a different number of unique classes, which leads to a complexity difference that we can leverage.
Each dataset consists of 60,000 data samples, where each dataset is split into 10,000 test data samples and 50,000 training samples.
In both cases, we use the predefined test data as server-side dataset for measuring the contribution.
The remaining data are partitioned with the Dirichlet distribution into non-IID datasets. 
We use this distribution as it is often used to create non-IID data~\cite{linEnsembleDistillationRobust2020,wangTacklingObjectiveInconsistency2020a}.
Here, we first sample the quantity for 10 participants and afterwards calculate the label distribution bounded to the quantity skew. 
Furthermore, we ensure that no sample is assigned to multiple participants. 
Hence, each participant has a unique label distribution and amount of data.
We calculate the skew for 10 participants in all cases, even if only 4 participants are in the process. 
This way, we can simulate that new participants introduce new data. 
We have two different skews that are created through varying alpha values of the Dirichlet distribution:
\begin{itemize}
    \item \textit{Base skew} consists of a quantity skew alpha of 2.5 and an alpha value of 1 for the label skew (see Figure~\ref{tab:bs_client_data}).
    \item \textit{High label skew} distributes the quantity more uniformly (alpha of 10), but the alpha for the label skew is 0.2.
\end{itemize}
For the CIFAR-10 dataset, that has a smaller number of classes, we use base skew and high label skew to analyze if different skews cause different behaviors.
In case of the CIFAR-100 dataset, we only use base skew as the dataset is already very complex.
\begin{table}[t]
\centering
\caption{CIFAR-10 Base Skew: Samples per label for each participant ordered by (0, 1, 2, 3, 4, 5, 6, 7, 8, 9)}
\label{tab:bs_client_data}
\begin{tabular}{|l|l|}
\hline
Participant & Samples per Label \\
\hline
0 & 454,  1604, 189, 243, 2737, 478, 939, 1473, 377, 413 \\
1 & 683,  287,  501, 417, 573, 170,  167, 261, 31,   396 \\
2 & 550,  162,  90,  208, 16,  361,  18,  183, 98,   315 \\
3 & 233,  1650, 832, 706, 66,  2435, 28,  141, 1435, 36 \\
\hline
4 & 58,   211,  666, 336, 118, 259,  945, 190, 89,   526 \\
5 & 1262, 203,  1633, 640, 413, 113, 301, 1502, 1694, 909 \\
\hline
6 & 1126, 326,  266, 1520, 497, 3,  1868, 818, 156,  1639 \\
7 & 31,   143,  121, 143, 72,  231, 58,  22,   756,  228 \\
\hline
8 & 23,   263,  219, 567, 352, 215, 341, 329,  56,   131 \\
9 & 575,  146,  479, 216, 153, 731, 332, 78,   305,  403 \\
\hline
\end{tabular}
\end{table}

\subsubsection{Model Architecture}
For both datasets, we use specific model architectures to achieve reasonable performance for each task.
In case of the CIFAR-10 dataset, we use the convolutional neural network (CNN) model described in Table~\ref{tab:cifar_10_model_arch}.
For the CIFAR-100 dataset we implement the model in Table~\ref{tab:cifar_100_model_arch}.
We choose this architecture as it achieves reasonable performance and also enables us to simulate the training of 10 participants.

\begin{table}[t!]
\centering
\caption{CIFAR-10: Model Architecture}
\label{tab:cifar_10_model_arch}
\begin{tabular}{|l|c|}
\hline
\textbf{Order} & \textbf{Layer} \\
\hline
1 & Conv2D(32, (3, 3), activation='relu', input\_shape=(32, 32, 3)) \\
2 & MaxPooling2D() \\
3 & Conv2D(64, (3, 3), activation='relu') \\
4 & MaxPooling2D()\\
5 & Conv2D(64, (3, 3), activation='relu') \\
6 & Flatten() \\
7 & Dense(64, activation='relu') \\
8 & Dense(10, activation='softmax') \\
\hline
\end{tabular}
\end{table}

We implemented both models with Tensorflow\footnote{https://www.tensorflow.org/}.
During the participant setup, each participant initially performs a train-test split with a ratio of 70-30.

\begin{table}[t!]
\centering
\caption{CIFAR-100: Model Architecture}
\label{tab:cifar_100_model_arch}
\begin{tabular}{|l|c|}
\hline
\textbf{Order} & \textbf{Layer} \\
\hline
1 & Conv2D(64, (3, 3), padding='same', input\_shape=(32, 32, 3))  \\
2-3 & BatchNormalization(), Activation('relu') \\
4 & Conv2D(64, (3, 3), padding='same') \\
5-6 & BatchNormalization(), Activation('relu') \\
7 & MaxPooling2D()\\
8 & Dropout(0.2, seed=24) \\
\hline
9-24 & 2x Repeat 1-8 \\
\hline
25 & Flatten() \\
26 & Dense(128, activation='relu') \\
27 & BatchNormalization()\\
28 & Dense(100, activation='softmax')\\
\hline
\end{tabular}
\end{table}

\begin{table}[t!]
\centering
\caption{Base Configurations for CIFAR-10 and CIFAR-100}
\label{tab:base_configs}
\begin{tabular}{|l|c|c|}
\hline
\textbf{Parameter} & \textbf{CIFAR-10} & \textbf{CIFAR-100} \\
\hline
Initial Number Of Participants & 4          & 4             \\
\hline
Quantity Skew Alpha            & 2.5        & 2.5           \\
Label Skew Alpha               & 1          & 1             \\
\hline
Dropout Participating Rounds   & 2          & 2             \\
Availability Timing            & Start      & Start         \\
\hline
Local Training Rounds          & 5 Rounds   & 5 Rounds      \\
FL Training Rounds             & 20 Rounds  & 20 Rounds     \\
Aggregation Strategy           & FedAvg     & FedAvg\\
Optimizer (Learning Rate)      & Adam(0.0001) & Adam(0.0001)\\
Metrics                        & Accuracy, F1 score & Accuracy, F1 \\
\hline
\end{tabular}
\end{table}

\subsubsection{Configurations}
In order to analyze the influence of the factors, we perform different experiments based on the number of participants, data skew, and availability timing.
For analyzing the impact of single factor, we use the base configuration and run multiple experiments while only adjusting the selected factor following Table~\ref{tab:configuration_range}.
In some cases, if the experiments show promising results, we perform additional experiments.

\begin{table}[t!]
\centering
\caption{Range of parameters for CIFAR-10 and CIFAR-100 }
\label{tab:configuration_range}
\begin{tabular}{|l|c|c|}
\hline
\textbf{Factors}           & \textbf{Range of parameters} \\
\hline
Number Of Participants     &  \textbf{4}, 6, 8, 10              \\
Data Skew (Quantity, Label)&  \textbf{(2.5,1)}, (10,0.2)     \\
Availability Timing        &  \textbf{Start}, Middle, End   \\
\hline
\end{tabular}
\end{table}

In our experimentation environment, we decide beforehand which participant drops out and only the dropout of one participant per FL process is considered.
We mostly consider that participant 0 drops out, as this one holds the most data. Therefore, we assume that this dropout has the highest impact.
Similarly, the missing participant is only active for 2 rounds and will miss the rest of the FL process.
The timing of the participation is at the start of the process (i.e, the first two rounds), in the middle of the process, or during the last two rounds.
This way we can analyze, concrete impacts that can happen if a participant joins and leaves at different times.
In our experiment, a participant failure makes the participant unable to participate in the process this includes training but also the evaluation.

For evaluation and comparison of the dropout impact, we use the same experiment, however, without a failing participant.
Furthermore, for every round, we perform participant-based evaluation and aggregate the evaluation metrics in two ways.
The first one with all participants and the second one excludes the missing participant.
While CIFAR-10 and CIFAR-100 are balanced datasets, the dropout can break the balance; hence, it is important to use the F1 score as this can handle both balanced and imbalanced datasets.
For consistent results, we use seeds in the experiments.

\section{Findings}
\label{sec:findings}
In this section, we will discuss the key findings from our experiments.

\begin{figure}[t]
    \centering
    \includegraphics[width=245pt]{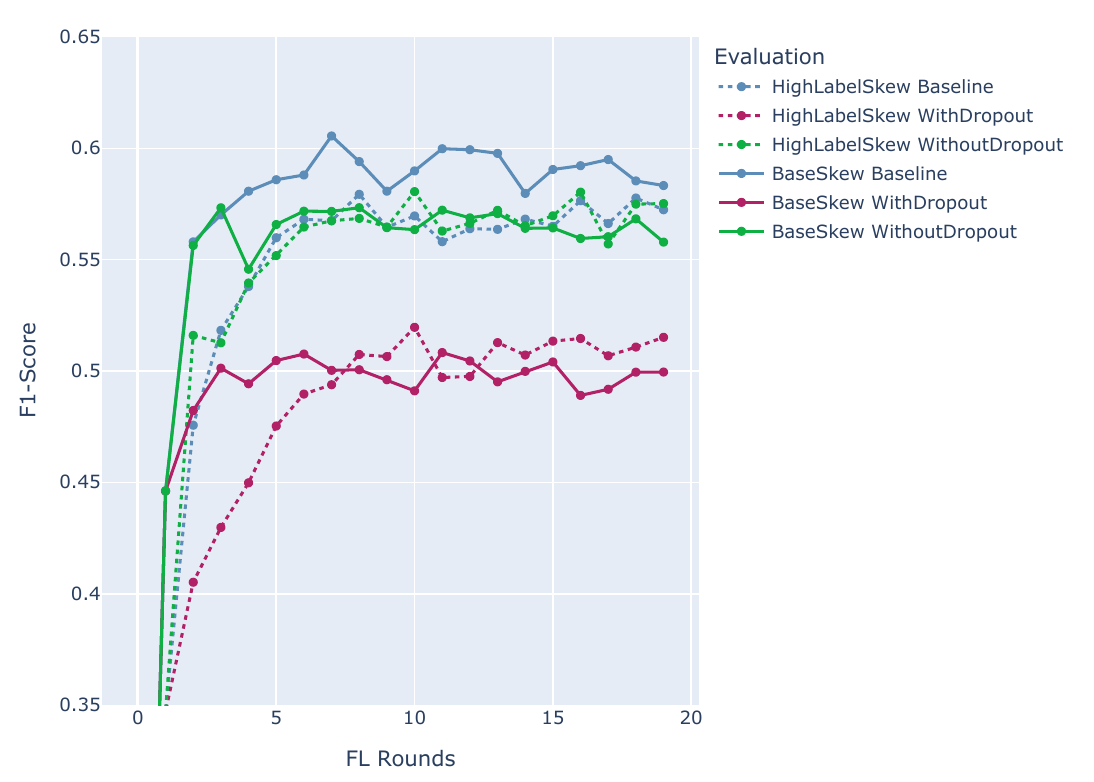}
    \caption{CIFAR-10: Comparison of the evaluation with and without the failed participant across different skews.}
    \label{fig:impact_evaluation}
\end{figure}

\noindent\textbf{Finding 1: A missing participant during evaluation leads to an optimistic evaluation for skewed datasets --}
The evaluation of ML models is crucial for assessing their performance.
In FL, the evaluation is done by the participants; hence, if a participant is missing this also impacts the evaluation.
The experiments showed, that for the CIFAR-10 dataset, the impact of the missing data during the evaluation, makes the evaluation biased towards higher quality.
Figure~\ref{fig:impact_evaluation} shows the difference between evaluation with and without the missing participant for the CIFAR-10 dataset. We can identify that in both skews, the evaluation excluding the missing participant leads to a higher F1 score than if all participants are included in the evaluation.

\begin{figure}[t]
    \centering
    \includegraphics[width=245pt]{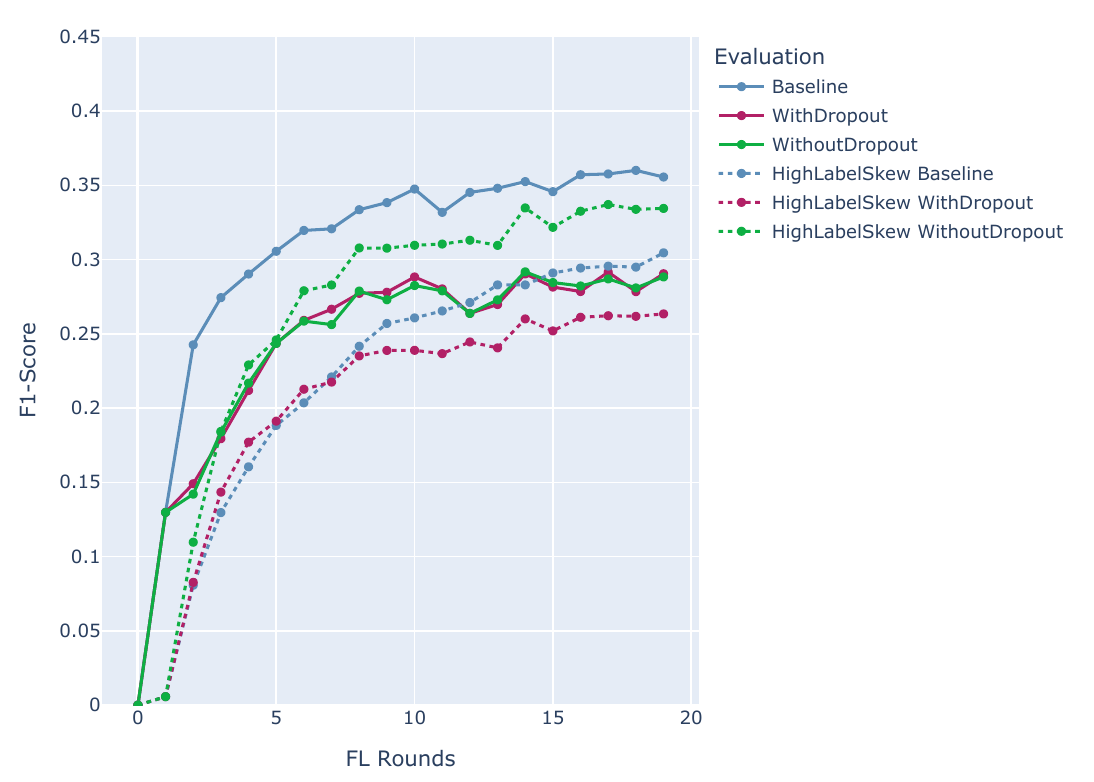}
    \caption{CIFAR-100: Comparison of the evaluation with and without the failed participant across different skews.}
    \label{fig:cifar_100_impact_evaluation}
\end{figure}

In comparison, with the results for CIFAR-100 with base skew in Figure~\ref{fig:cifar_100_impact_evaluation}, the impact of the missing participant during evaluation is similar to including all participants.
We assume that the important factors that cause this behavior are the generalization layers of the model, the high number of classes and the data skew. 
The reason is that while participant 0 has the largest amount of data, the number of classes is high and therefore, there are only a few samples per class. 
The base skew configuration is generating a quantity skew with less focus on the label skew; hence, each participant has samples for each class.
Since the model includes multiple generalization layers, the individual impact of the quantity per class is low as the model tries to generalize over all classes.
Therefore, during the evaluation, a large quantity of a specific class is less important.
Hence, the impact of the missing participant on the evaluation is low if the classes are still represented in the test set and only the number is reduced. 

We further support this assumption by comparing the combined label and quantity distribution of all participants with the combined distribution that excludes participant 0.
The overall samples for testing decreased from 4311 to 1850 and each class is represented in the test dataset, with the lowest number of samples per class being 3 and the highest being 47. 
In comparison, with all participants included, the highest number of samples per class is 116 and the lowest is 6.
We conclude that with the missing participant, the overall dataset for testing is only reduced in quantity; hence, the difference between both evaluations is similar.
Furthermore, to support this assumption, we analyze the high label skew configuration for CIFAR-100 in Figure~\ref{fig:cifar_100_impact_evaluation}, which shows the overly optimistic evaluation once labels are highly skewed.
Consequently, if a participant drops out during evaluation the testing goal can deviate.
This finding is very important as there are many methods that build upon utility measurement (e.g., early stopping) and, hence, if a participant failure occurs this can lead to unexpected behavior which affects the reliability of a system.

\begin{figure}[t]
    \centering
    \includegraphics[width=235pt]{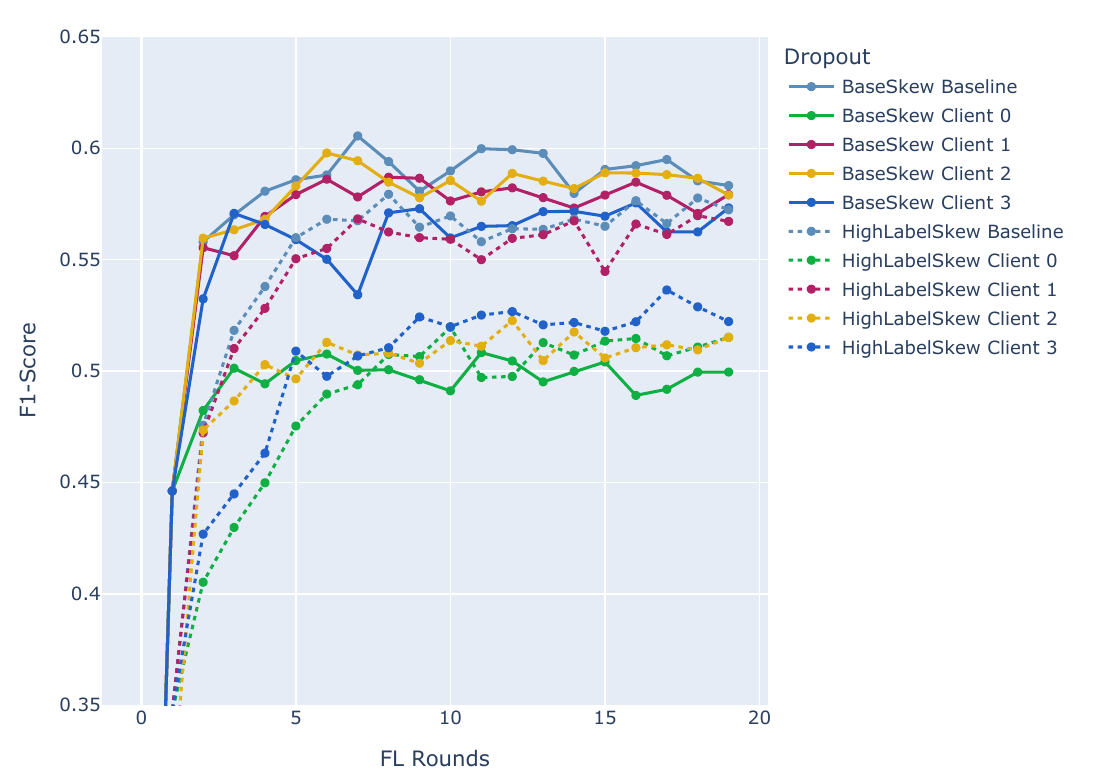}
    \caption{CIFAR-10: Impact of different participants dropping out on the performance across different skews.}
    \label{fig:impact_of_different_participant_dropping_out}
\end{figure}

\noindent\textbf{Finding 2: Participant contribution metrics can indicate which participant dropout would have the highest impact on the model performance --}
For this finding, we analyze the impact of participant failures at a specific round on the global model.
The participants in our base configuration fail and dropout after the second round. It is important to understand, which SVs the participants have up to the point before the dropout occurs.
For CIFAR-10 with the base skew, Table~\ref{tab:cifar_10_shapley_values_hs} shows the SVs.
For base skew, the order is 0, 3, 2, and 1.
If we compare the SVs with the impacts of different participants that drop out (see Figure~\ref{fig:impact_of_different_participant_dropping_out}), one can identify that the missing participant 0 has the highest impact and the second value has the second-highest impact.
The smaller values show a roughly similar impact.
Next, we analyze the SVs for the high label skew in Table~\ref{tab:cifar_10_shapley_values_hs}.
Here, they indicate that participant 0 has the highest SV followed by 1, 2, and 3.
If we compare it with the impact of the individual dropouts in Figure~\ref{fig:impact_of_different_participant_dropping_out}, we can see that participant 0 has the highest immediate impact and at the last round the impact is similar to participant 2.
In this context, the SVs are only able to indicate the impact for the most valuable participant.
\begin{table}[t!]
\centering
\caption{CIFAR-10: Shapley values for base skew}
\label{tab:cifar_10_shapley_values}
\begin{tabular}{|l|c|c|c|c|}
\hline
\textbf{Round} & \textbf{P0-SV} & \textbf{P1-SV} & \textbf{P2-SV2}  & \textbf{P3-SV}  \\
\hline
0 &	0.036014&	-0.013779 & -0.014793 & 0.009743\\
1 & 0.066771 &	-0.013143 & -0.036243 & 0.004950\\
\hline
\end{tabular}
\end{table}

\begin{table}[t!]
\centering
\caption{CIFAR-10: Shapley values for high label skew}
\label{tab:cifar_10_shapley_values_hs}
\begin{tabular}{|l|c|c|c|c|}
\hline
\textbf{Round} & \textbf{P0-SV} & \textbf{P1-SV} & \textbf{P2-SV2}  & \textbf{P3-SV}  \\
\hline
0 &	0.069279 &	0.006957 & -0.036521 & -0.029793\\
1 & 0.104086 &	0.054114 & -0.055886 & -0.084857\\
\hline
\end{tabular}
\end{table}

\begin{figure}[t]
    \centering
    \includegraphics[width=235pt]{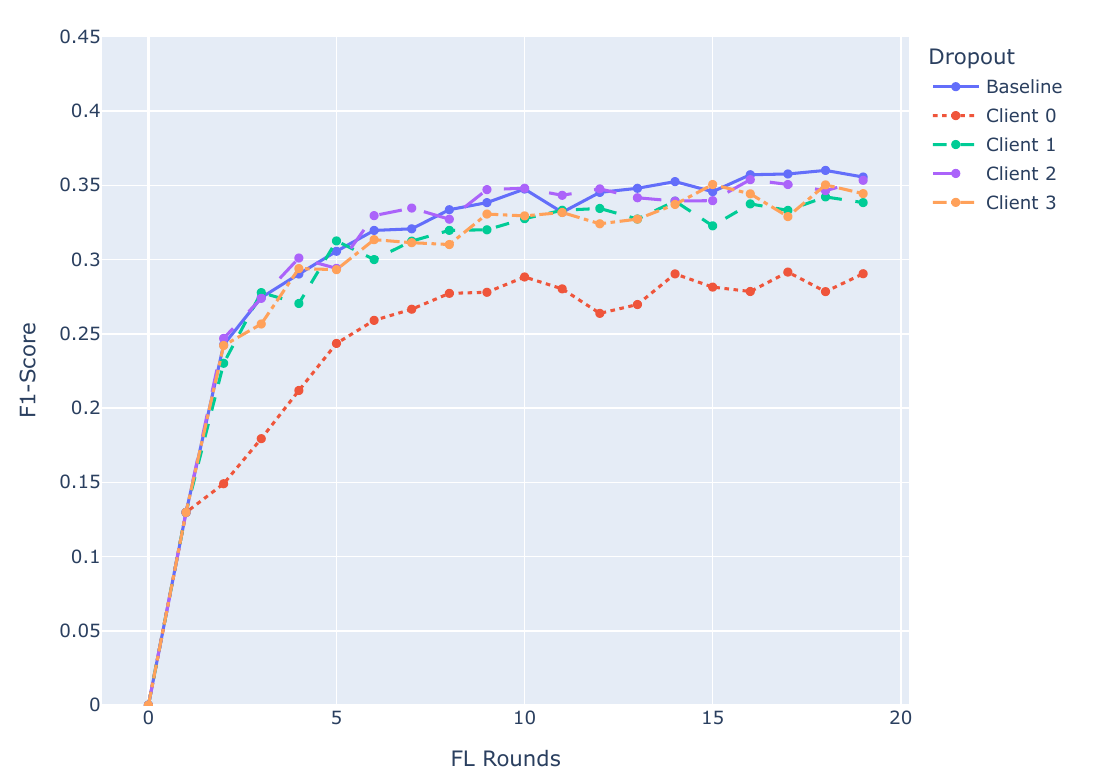}
    \caption{CIFAR-100: Impact of different participants dropping out on the performance.}
    \label{fig:cifar_100_impact_of_different_participant_dropping_out}
\end{figure}

\begin{table}[t!]
\centering
\caption{CIFAR-100: Shapley values for base skew}
\label{tab:cifar_100_shapley_values}
\begin{tabular}{|l|c|c|c|c|}
\hline
\textbf{Round} & \textbf{P0-SV} & \textbf{P1-SV} & \textbf{P2-SV2}  & \textbf{P3-SV}  \\
\hline
0 &	0.036836 &	-0.009286 & -0.016186 & -0.010850 \\
1 & 0.055921 &	-0.013886 & -0.017757 & -0.021793 \\
\hline
\end{tabular}
\end{table}
We also analyze the SVs for CIFAR-100 (see Table~\ref{tab:cifar_100_shapley_values}).
Similar to the CIFAR-10 base skew configuration, participant 0 has the highest value, followed by 1, 2, and 3. 
By comparing this order with the impact on the performance (Figure~\ref{fig:cifar_100_impact_of_different_participant_dropping_out}), one can observe that the impact for the highest valued participant and the second-highest values is correct.
However, the SVs for participant 1, 2, and 3 only have small differences similar to their impact.
Since, the method is based on estimations we expect some (low) variability.
Consequently, the SVs provide a good indication about the immediate impact of a high value participant.
This finding is important as it can help in deciding on a suitable fault-tolerance mechanisms for mitigating participant failures.

\begin{figure}[t]
    \centering
    \includegraphics[width=235pt]{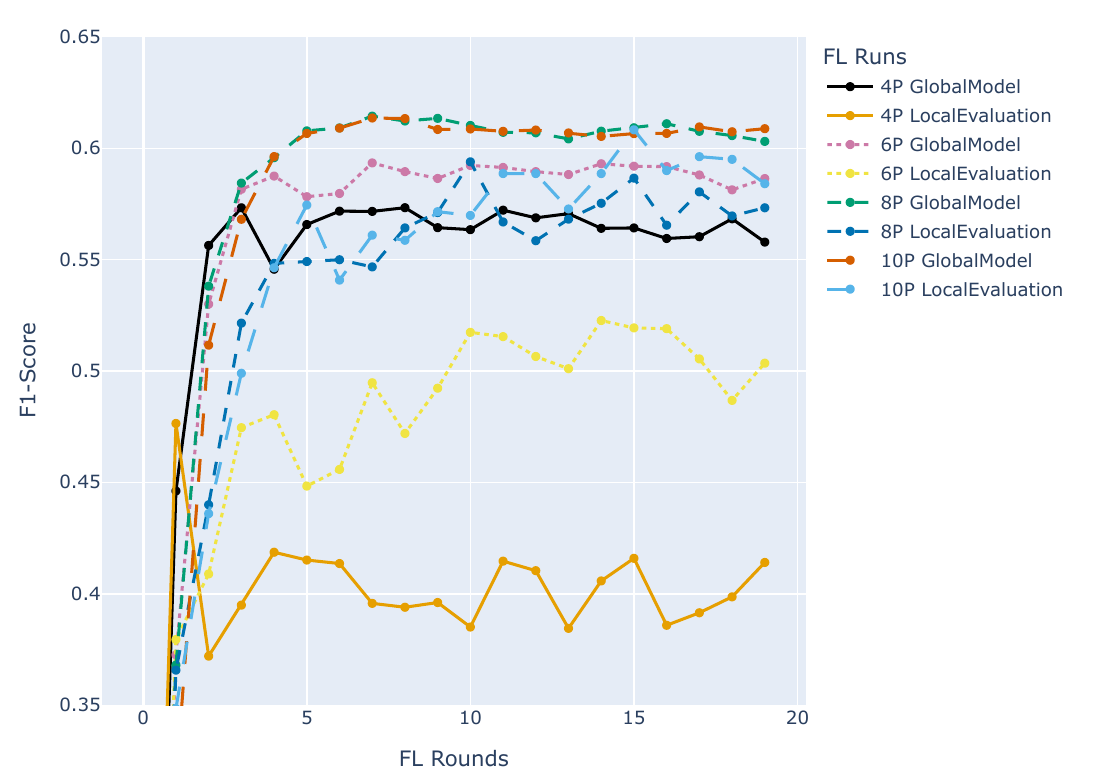}
    \caption{CIFAR-10: Comparison of the global model's performance for the failed participant against the performance for the other participants.}
    \label{fig:impact_local_vs_global_performance}
\end{figure}

\noindent\textbf{Finding 3: The failed participant can use the global model with little performance loss for its task without additional effort if other participants have similar information --}
If a participant drops out of the process depending on the behavior of the dropout, its information may not be included and can be forgotten by the model over time; however, the participant also wants to use the trained model. Therefore, it is important to understand how the participant failure affects the applicability of the global model for the missing participant.
In Figure~\ref{fig:impact_local_vs_global_performance}, we compare the local evaluation of the global model for the failed participant against the evaluation of the global model where the missing participant is excluded.
We exclude the evaluation metrics, because otherwise the evaluation would be biased.
The experiments for CIFAR-10 show that the number of participants impact the quality of the global model: with more participants the performance improves as relevant information is added to the FL process.

Furthermore, one can see that the performance is significantly lower for the failed participant in a scenario with 4 participants compared to the global model performance with 4 participants. 
In addition, the more participants of the federation possess similar information, the lower is the performance loss of the resulting model for a failed participant.
One can assume that the similarity of data leads to the substitution of lost information by the remaining participants.

\begin{figure}[t]
    \centering
    \includegraphics[width=235pt]{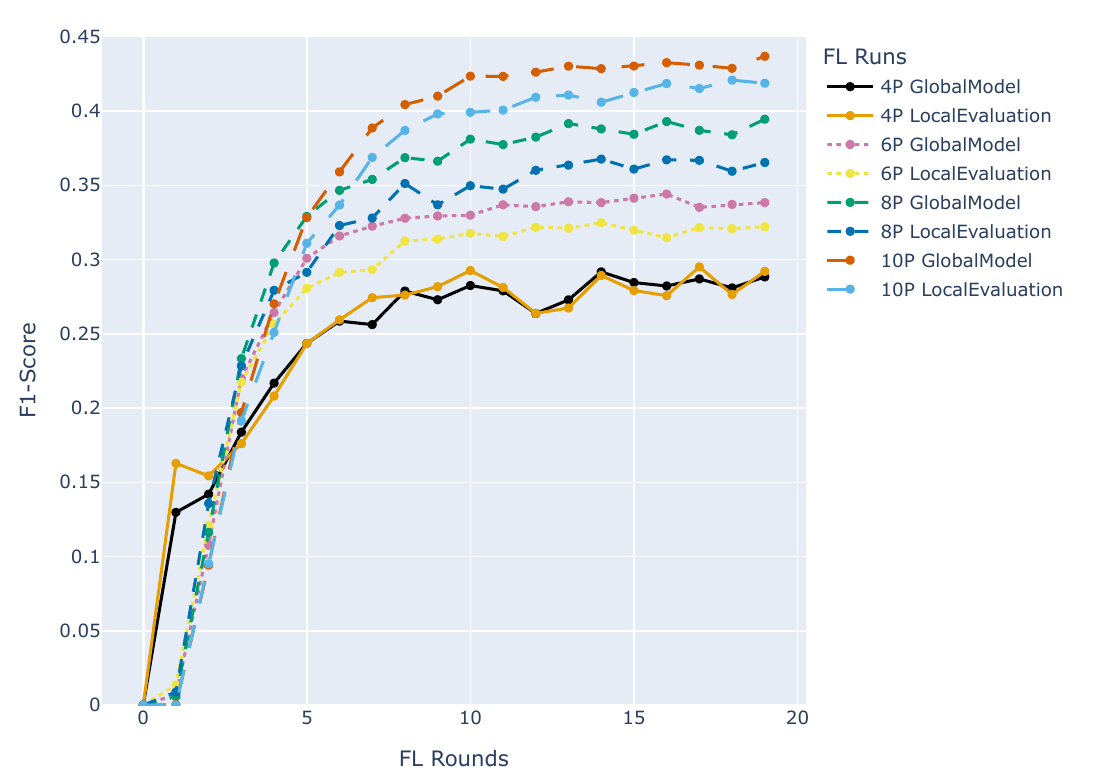}
    \caption{CIFAR-100: Comparison of the global model's performance for the failed participant against the performance for the other participants.}
    \label{fig:impact_local_vs_global_performance_cifar100}
\end{figure}

Next, for the CIFAR-100 dataset in Figure~\ref{fig:impact_local_vs_global_performance_cifar100}, more participants also improve the overall quality of the global model.
For 4 participants, we have the same behavior as in the evaluation comparison, which is caused by the number of classes, the amount of data, and the generalization aspects. Here, the performance loss is negligibly low.
We can identify that the difference for 6 participants increases and is the highest for 8 participants. 
For 10 participants, the difference is getting smaller compared to 8 participants.
Upon analyzing the information that is added by new participants, we identified that the new information was leading to higher skew and upon reaching 8 participants, it peaked. 
After reaching 10 participants, the full CIFAR-100 dataset is used; hence, leading to balanced samples per label in the FL process.
We identified a similar behavior for the CIFAR-10 dataset with high label skew and also inspected the different label distributions.
The data introduced by the configuration with 6 participants contained each label with more balanced samples per labels, hence, introducing samples for each class. 
Contrary, for 8 participants and 10 participants, the added information was highly skewed.
However, upon reaching 10 participants the full CIFAR-10 dataset is used; hence, the difference is getting smaller.
If more similar data are introduced to the process, the impact of the dropout can be reduced and the applicability of the global model for the dropout improves.
We conclude that it is very important to understand and analyze the information introduced to the FL process by each participant.

\noindent\textbf{Finding 4: The robustness against participant failures can be improved if the number of participants with similar data increase --}
In the previous paragraph, both Figure~\ref{fig:impact_local_vs_global_performance} and Figure~\ref{fig:impact_local_vs_global_performance_cifar100} showed, that with more participants adding information to the process, the robustness against dropouts improves.
However, this is only the case if the new participant has similar data matching the current data distribution of the whole process.
Therefore, information can be substituted after failure, as the experiment showed.
However, if we assume that a new participant introduces new unique classes, then it is unlikely that it can supplement information from other participants.
In this regard, it is important to consider data heterogeneity.

\begin{figure}[t]
    \centering
    \includegraphics[width=235pt]{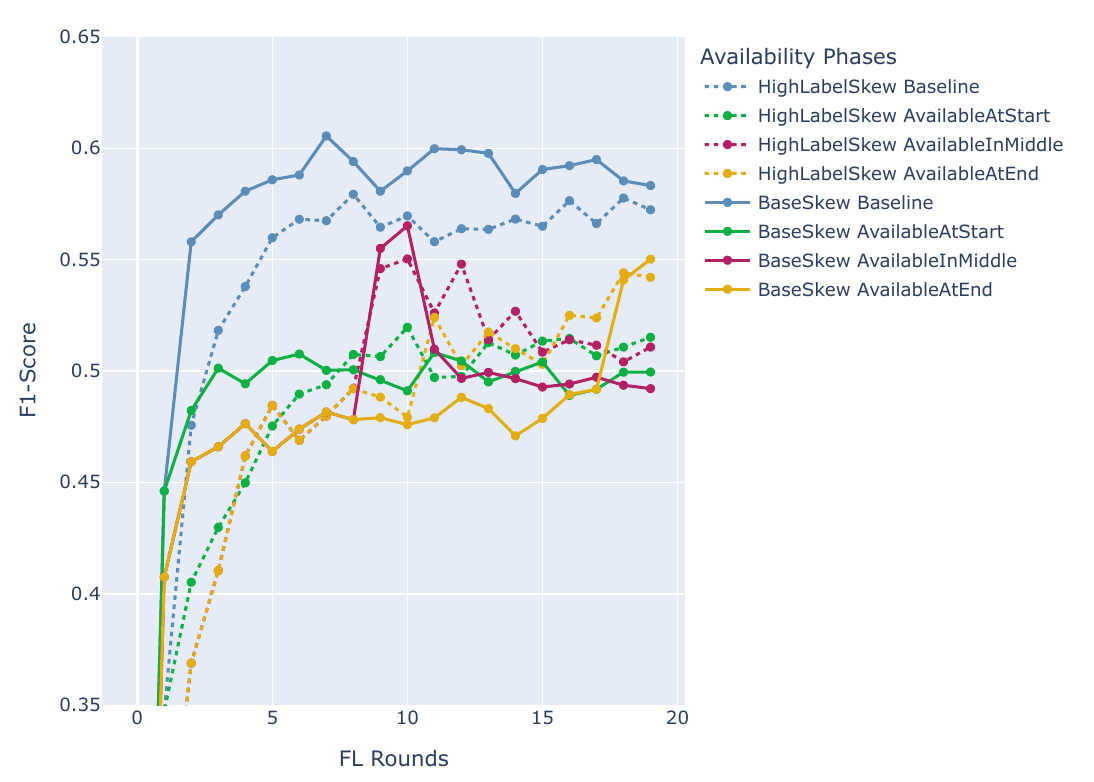}
    \caption{CIFAR-10: Impact of different availability phases on the performance of the global model across different skews.}
    \label{fig:availability_impact_cifar10}
\end{figure}

\begin{figure}[t]
    \centering
    \includegraphics[width=235pt]{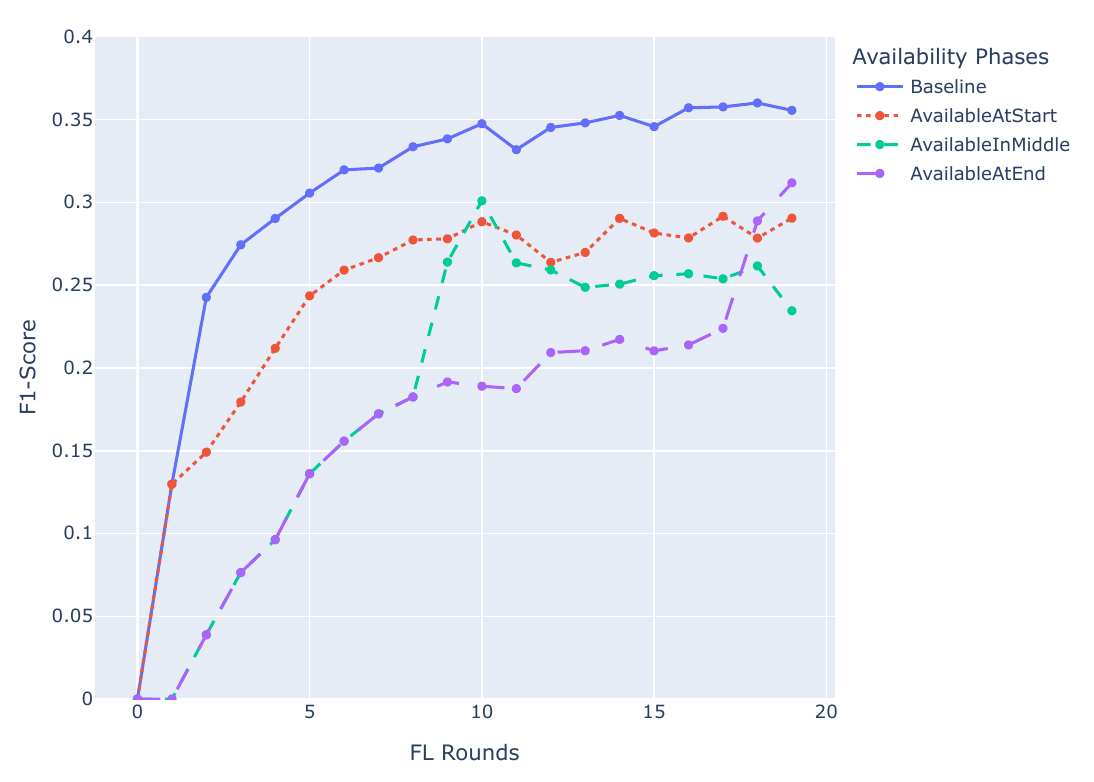}
    \caption{CIFAR-100: Impact of different availability phases on the performance of the global model.}
    \label{fig:availability_impact_cifar100}
\end{figure}

\noindent\textbf{Finding 5: The timing of the dropout affects the quality of the trained model --}
Next, we analyze the impact of participants' availability.
In Figure~\ref{fig:availability_impact_cifar10} the base skew and the high label skew are visualized for the CIFAR-10 dataset.
One can see that in each availability phase the performance spikes because the model learns additional information. 
However, this spike only lasts for the duration of the participation and drops over the next few rounds to the previous performance.
In this regard, the highest impact on the global model was given, where participant 0 drops out of the whole process and re-joins for the last two rounds.
Within this timeframe, we assume that the model has learned new patterns improving the quality of the global model.
Furthermore, this behavior remains consistent for both skew configurations.
If we compare the results to CIFAR-100 (Figure~\ref{fig:availability_impact_cifar100}), then this further supports the previous results. 
The immediate impact causes the model to receive additional information, which can help in learning patterns.
In addition, the participation availability in the middle of the FL process highlights that the model learns information, retains them, and then forgets the information after a few rounds.
We assume that the impact of forgetting is higher in FL as compared to centralized ML because, beyond the number of local training rounds and learning rate, the aggregation method (i.e., FedAvg~\cite{mcmahanCommunicationEfficientLearningDeep2017a}) strongly influences the change of model weights.
This finding is important because it highlights the impact of participants being available early or joining later to the process.
Therefore, it can help in making decisions to decide how to handle a dropout (e.g., a participant failure occurs early but recovers later).

\subsection{Validity of the Findings} 

\label{sec:validity}

In this section, we critically discuss aspects of the validity of our experimentation study.

Firstly, regarding the use of SVs, it can be argued that these are accurate and allow for analyzing, if participant contribution metrics can indicate the impact of a failed participant.
Nevertheless, SV computation requires a dataset to be available at the FL coordinator, which is not feasible for most real-world applications.
In order to make contribution metrics also available as an indicator for supporting failure handling, future work should analyze alternative metrics that focus on real-world applicability.
Secondly, we have focused a selected range of factors, which did not include the impact of the model architecture or state-of-the-art techniques such as data augmentation to achieve the highest possible performance. 
We assume that a larger model architecture will lead to improved baseline performance but also to a higher failure impact on the performance as the data do not change and the training requires more time.
In addition, we considered that one participant fails and drops out of the FL process. 
While this assumption is helpful for analyzing the impact of specific participants failing, in a real-world setting, multiple participants might fail at the same time. 
With an increasing number of failing participants, the data available for training and evaluation will be reduced further. 
Therefore, we assume that multiple missing participants will lead to a higher impact on the quality.

\section{Conclusion}
% Conclusion
In cross-silo FL with few participants, it is important to consider and assess potential risks such as the failure of a participant in order to provide a reliable model.
However, the impact of a dropout is mostly studied in cross-device FL and is missing a detailed analysis for cross-silo FL.

In this paper, we conducted an experimentation study % konsistente Terminologie
that analyzes the impact of a missing participant on the model performance while considering a subset of influential factors (i.e., data complexity, evaluation, and timing of the dropout).
We showed that a Shapley value-based participant contribution metric can be used for indicating the immediate impact of high-value participants.
Furthermore, we showed that the timing and the data of the failed participant influences the quality of the trained model significantly.
Additionally, if the dropped out participant is also missing for the model evaluation then this can result in an overly optimistic evaluation.
We assume that our findings can help future work by understanding the impact of participant dropouts in cross-silo FL, while also providing crucial background for the design of failure handling methods.

\bibliographystyle{IEEEtran}

\bibliography{bibliography}

@article{bonawitzFederatedLearningScale2019,
  title = {Towards {{Federated Learning}} at {{Scale}}: {{System Design}}},
  shorttitle = {Towards {{Federated Learning}} at {{Scale}}},
  author = {Bonawitz, Keith and Eichner, Hubert and Grieskamp, Wolfgang and Huba, Dzmitry and Ingerman, Alex and Ivanov, Vladimir and Kiddon, Chlo{\'e} and Kone{\v c}n{\'y}, Jakub and Mazzocchi, Stefano and McMahan, Brendan and Van Overveldt, Timon and Petrou, David and Ramage, Daniel and Roselander, Jason},
  year = {2019},
  month = apr,
  journal = {Proceedings of Machine Learning and Systems},
  volume = {1},
  pages = {374--388},
  urldate = {2023-06-02},
  langid = {english}
}

@inproceedings{paper_stricker_fl-apu,
  title={FL-APU: A Software Architecture to Ease Practical Implementation of Cross-Silo Federated Learning},
  author={Stricker, Fabian and Perez, Jose Antonio Peregrina and Zirpins, Christian and Bermbach, David},
  booktitle={Proceedings of the 2nd International Conference on Federated Learning Technologies and Applications (FLTA 2024)},
  year={2025},
  organization={IEEE}
}

@inproceedings{ghorbaniDataShapleyEquitable2019,
  title = {Data {{Shapley}}: {{Equitable Valuation}} of {{Data}} for {{Machine Learning}}},
  shorttitle = {Data {{Shapley}}},
  booktitle = {Proceedings of the 36th {{International Conference}} on {{Machine Learning}}},
  author = {Ghorbani, Amirata and Zou, James},
  year = {2019},
  pages = {2242--2251},
  publisher = {PMLR},
  issn = {2640-3498},
  urldate = {2025-05-31},
  langid = {english}
}

@misc{huangKeepItSimple2023,
  title = {Keep {{It Simple}}: {{Fault Tolerance Evaluation}} of {{Federated Learning}} with {{Unreliable Clients}}},
  shorttitle = {Keep {{It Simple}}},
  author = {Huang, Victoria and Sohail, Shaleeza and Mayo, Michael and Botran, Tania Lorido and Rodrigues, Mark and Anderson, Chris and Ooi, Melanie},
  year = {2023},
  eprint = {2305.09856},
  primaryclass = {cs},
  doi = {10.1109/CLOUD60044.2023.00024},
  urldate = {2025-04-02},
  archiveprefix = {arXiv}
}

@article{imteajSurveyFederatedLearning2022,
  title = {A {{Survey}} on {{Federated Learning}} for {{Resource-Constrained IoT Devices}}},
  author = {Imteaj, Ahmed and Thakker, Urmish and Wang, Shiqiang and Li, Jian and Amini, M. Hadi},
  year = {2022},
  journal = {IEEE Internet of Things Journal},
  volume = {9},
  number = {1},
  pages = {1--24},
  issn = {2327-4662},
  doi = {10.1109/JIOT.2021.3095077},
  urldate = {2025-06-21}
}

@article{kairouzAdvancesOpenProblems2021,
  title = {Advances and Open Problems in Federated Learning},
  author = {Kairouz, Peter and McMahan, H Brendan and Avent, Brendan and Bellet, Aur{\'e}lien and Bennis, Mehdi and Bhagoji, Arjun Nitin and Bonawitz, Kallista and Charles, Zachary and Cormode, Graham and Cummings, Rachel and others},
  year = {2021},
  journal = {Foundations and trends{\textregistered} in machine learning},
  volume = {14},
  number = {1--2},
  pages = {1--210}
}

@inproceedings{kingmaAdamMethodStochastic2015,
  title = {Adam: {{A Method}} for {{Stochastic Optimization}}},
  shorttitle = {Adam},
  booktitle = {International {{Conference}} on {{Learning Representations}} ({{ICLR}})},
  author = {Kingma, Diederik P. and Ba, Jimmy},
  year = {2015},
  eprint = {1412.6980},
  primaryclass = {cs},
  doi = {10.48550/arXiv.1412.6980},
  urldate = {2025-06-21},
  archiveprefix = {arXiv}
}

@article{krizhevskyLearningMultipleLayers2009a,
  title = {Learning {{Multiple Layers}} of {{Features}} from {{Tiny Images}}},
  author = {Krizhevsky, Alex},
  year = {2009},
  langid = {english}
}

@inproceedings{linEnsembleDistillationRobust2020,
  title = {Ensemble {{Distillation}} for {{Robust Model Fusion}} in {{Federated Learning}}},
  booktitle = {Advances in {{Neural Information Processing Systems}}},
  author = {Lin, Tao and Kong, Lingjing and Stich, Sebastian U and Jaggi, Martin},
  year = {2020},
  volume = {33},
  pages = {2351--2363},
  publisher = {Curran Associates, Inc.},
  urldate = {2025-06-26}
}

@article{liuGTGShapleyEfficient2022,
  title = {{{GTG-Shapley}}: {{Efficient}} and {{Accurate Participant Contribution Evaluation}} in {{Federated Learning}}},
  shorttitle = {{{GTG-Shapley}}},
  author = {Liu, Zelei and Chen, Yuanyuan and Yu, Han and Liu, Yang and Cui, Lizhen},
  year = {2022},
  journal = {ACM Trans. Intell. Syst. Technol.},
  volume = {13},
  number = {4},
  pages = {60:1--60:21},
  issn = {2157-6904},
  doi = {10.1145/3501811},
  urldate = {2025-06-21}
}

@misc{liuOutOfDistributionGeneralizationSurvey2023a,
  title = {Towards {{Out-Of-Distribution Generalization}}: {{A Survey}}},
  shorttitle = {Towards {{Out-Of-Distribution Generalization}}},
  author = {Liu, Jiashuo and Shen, Zheyan and He, Yue and Zhang, Xingxuan and Xu, Renzhe and Yu, Han and Cui, Peng},
  year = {2023},
  number = {arXiv:2108.13624},
  eprint = {2108.13624},
  primaryclass = {cs},
  publisher = {arXiv},
  doi = {10.48550/arXiv.2108.13624},
  urldate = {2025-06-21},
  archiveprefix = {arXiv}
}

@inproceedings{mcmahanCommunicationEfficientLearningDeep2017a,
  title = {Communication-{{Efficient Learning}} of {{Deep Networks}} from {{Decentralized Data}}},
  booktitle = {Proceedings of the 20th {{International Conference}} on {{Artificial Intelligence}} and {{Statistics}}},
  author = {McMahan, Brendan and Moore, Eider and Ramage, Daniel and Hampson, Seth and y Arcas, Blaise Aguera},
  year = {2017},
  pages = {1273--1282},
  publisher = {PMLR},
  issn = {2640-3498},
  urldate = {2025-06-20},
  langid = {english}
}

@inproceedings{qianDROPFLClientDropout2024a,
  title = {{{DROPFL}}: {{Client Dropout Attacks Against Federated Learning Under Communication Constraints}}},
  shorttitle = {{{DROPFL}}},
  booktitle = {2024 {{IEEE International Conference}} on {{Acoustics}}, {{Speech}} and {{Signal Processing}} ({{ICASSP}})},
  author = {Qian, Wenjun and Shen, Qingni and Xu, Haoran and Huang, Xi and Wu, Zhonghai},
  year = {2024},
  pages = {4870--4874},
  issn = {2379-190X},
  doi = {10.1109/ICASSP48485.2024.10446609},
  urldate = {2025-06-20}
}

@article{riberoFederatedLearningIntermittent2023a,
  title = {Federated {{Learning Under Intermittent Client Availability}} and {{Time-Varying Communication Constraints}}},
  author = {Ribero, Monica and Vikalo, Haris and Veciana, Gustavo De},
  year = {2023},
  journal = {IEEE Journal of Selected Topics in Signal Processing},
  volume = {17},
  number = {1},
  eprint = {2205.06730},
  primaryclass = {cs},
  pages = {98--111},
  issn = {1932-4553, 1941-0484},
  doi = {10.1109/JSTSP.2022.3224590},
  urldate = {2025-05-26},
  archiveprefix = {arXiv},
  langid = {english}
}

@article{sousaEnhancingRobustnessFederated2025,
  title = {Enhancing Robustness in Federated Learning Using Minimal Repair and Dynamic Adaptation in a Scenario with Client Failures},
  author = {Sousa, John and Ribeiro, Eduardo and Bustincio, Romulo and Bastos, Lucas and Morais, Renan and Cerqueira, Eduardo and Ros{\'a}rio, Denis},
  year = {2025},
  journal = {Annals of Telecommunications},
  issn = {1958-9395},
  doi = {10.1007/s12243-025-01075-3},
  urldate = {2025-05-26},
  langid = {english}
}

@article{sunMimiCCombatingClient2024,
  title = {{{MimiC}}: {{Combating Client Dropouts}} in {{Federated Learning}} by {{Mimicking Central Updates}}},
  shorttitle = {{{MimiC}}},
  author = {Sun, Yuchang and Mao, Yuyi and Zhang, Jun},
  year = {2024},
  journal = {IEEE Transactions on Mobile Computing},
  volume = {23},
  number = {7},
  pages = {7572--7584},
  issn = {1558-0660},
  doi = {10.1109/TMC.2023.3338021},
  urldate = {2025-05-26}
}

@inproceedings{wangFriendsHelpSaving2024,
  title = {Friends to {{Help}}: {{Saving Federated Learning}} from {{Client Dropout}}},
  shorttitle = {Friends to {{Help}}},
  booktitle = {2024 {{IEEE International Conference}} on {{Acoustics}}, {{Speech}} and {{Signal Processing}} ({{ICASSP}})},
  author = {Wang, Heqiang and Xu, Jie},
  year = {2024},
  pages = {8896--8900},
  issn = {2379-190X},
  doi = {10.1109/ICASSP48485.2024.10447268},
  urldate = {2025-02-06}
}

@incollection{wangPrincipledApproachData2020,
  title = {A Principled Approach to Data Valuation for Federated Learning},
  booktitle = {Federated Learning: {{Privacy}} and Incentive},
  author = {Wang, Tianhao and Rausch, Johannes and Zhang, Ce and Jia, Ruoxi and Song, Dawn},
  year = {2020},
  pages = {153--167},
  publisher = {Springer International Publishing},
  address = {Cham},
  isbn = {978-3-030-63076-8}
}

@inproceedings{wangTacklingObjectiveInconsistency2020a,
  title = {Tackling the {{Objective Inconsistency Problem}} in {{Heterogeneous Federated Optimization}}},
  booktitle = {Advances in {{Neural Information Processing Systems}}},
  author = {Wang, Jianyu and Liu, Qinghua and Liang, Hao and Joshi, Gauri and Poor, H. Vincent},
  year = {2020},
  volume = {33},
  pages = {7611--7623},
  publisher = {Curran Associates, Inc.},
  urldate = {2025-06-26}
}

@article{xuStabilizingImprovingFederated2025,
  title = {Stabilizing and Improving Federated Learning with Highly Non-Iid Data and Client Dropout},
  author = {Xu, Jian and Yang, Meilin and Ding, Wenbo and Huang, Shao-Lun},
  year = {2025},
  journal = {Applied Intelligence},
  volume = {55},
  number = {3},
  pages = {1--18}
}

@article{zhaoFederatedLearningNonIID2018,
  title = {Federated {{Learning}} with {{Non-IID Data}}},
  author = {Zhao, Yue and Li, Meng and Lai, Liangzhen and Suda, Naveen and Civin, Damon and Chandra, Vikas},
  year = {2018},
  eprint = {1806.00582},
  primaryclass = {cs, stat},
  doi = {10.48550/arXiv.1806.00582},
  urldate = {2023-04-17},
  archiveprefix = {arXiv}
}

\end{document}